\documentclass[12pt, a4paper]{article}
\usepackage{physics}
\usepackage{jheppub}
\usepackage{amsmath}
\usepackage{amssymb}
\usepackage{amsthm}
\usepackage{lmodern}
\usepackage{xcolor}
\usepackage{url}
\usepackage{cleveref}
\usepackage{hyperref}
\usepackage{graphicx}

\usepackage{chngcntr}
\counterwithout{equation}{section}

\usepackage[utf8]{inputenc}
\usepackage[T1]{fontenc}
\hypersetup{
  colorlinks=true,
  citecolor=magenta,
  linkcolor=blue,
  urlcolor=violet
 }

\newcommand{\del}{\partial}
\newcommand{\mi}{\mathrm{i}}
\def\be{\begin{equation}}\def\ee{\end{equation}}
\newcommand{\baa}{\begin{equation}\begin{aligned}}
\newcommand{\ea}{\end{aligned}\end{equation}}

\title{Dirac-Born-Infeld Counter-Term and \\Black Hole Thermodynamics}

\author[\dag\ddag]{Dileep P. Jatkar}
\author[*]{and Upamanyu Moitra}

\affiliation[\dag]{Harish-Chandra Research Institute\\ Chhatnag Road, Jhunsi, Allahabad 211019,  India}

\affiliation[\ddag]{Homi Bhabha National Institute\\ Training School Complex, Anushaktinagar\\ Mumbai 400094, India}

\affiliation[*]{International Centre for Theoretical Physics\\
Strada Costiera 11,  Trieste 34151, Italy}

\emailAdd{dileep@hri.res.in}
\emailAdd{umoitra@ictp.it}

\abstract{We revisit the Dirac-Born-Infeld--like boundary counter-term
  for four dimensional theory of gravity.  We show that it correctly
  executes complete background subtraction for both asymptotically AdS
  and asymptotically flat geometries.  With an appropriate choice of
  ensemble, we reproduce dyonic black hole thermodynamics with both
  types of asymptotics by studying local thermodynamics on the cut-off
  surface.}
\makeatletter
\gdef\@fpheader{ \\ }
\makeatother
\begin{document}

\maketitle

\section{Introduction}

In order to develop a better understanding of dualities in string
theory, a lot of ideas are being pursued in comparing information
theoretic measures between dual theories.  One of the prime examples
of dualities which provides a non-gravitational dual description of
gravity is the AdS/CFT correspondence.  As is well-known, this
correspondence gives a holographic dictionary between the boundary
gauge theory and bulk gravitational theory.  The
entanglement entropy of two sub-regions is one of the information
theoretic measures which has been the focus of these investigations.
These computations within quantum field theory are plagued with
ultraviolet divergences but they are particularly hard within the
gauge theory context due to gauge redundancies.  The holographic dual
within the gauge/gravity correspondence was proposed by Ryu and
Takayanagi (RT) \cite{Ryu:2006ef} in terms of the area of the minimal
surface which bounds the entangling surface in the gauge theory.  Its
covariant generalisation was proposed in \cite{Hubeny:2007xt}.  The
minimal area surface typically dips into the bulk geometry where
gravitational theory is in charge of the affairs.

The ultraviolet (UV) divergences that appear in the quantum field
theory computation have bulk analogues in the form of infrared (IR)
divergences and one needs to take care of them by introducing boundary
counter-terms, which are over and above the Gibbons-Hawking-York term
\cite{York:1972sj, Gibbons:1976ue} that is needed to ensure that the
variational principle is well defined.  Balasubramanian and Kraus (BK)
\cite{Balasubramanian:1999re} proposed a class of counter-terms that
take care of divergent terms and give finite answers when we remove
the ultraviolet cut-off in the boundary quantum field theory, which
corresponds to taking the bulk radial cut-off to infinity.  While this
serves the purpose for computations of correlation functions in the
boundary theory,  or equivalently, computation of Witten diagrams in
the bulk gravitational theory,  this may not be sufficient when one is
looking at extended objects like the minimal area surface which probes
multiple radial foliations of the bulk geometry.  Holographic
renormalisation \cite{Henningson:1998gx, Emparan:1999pm,
  Kraus:1999di,Skenderis:2002wp} is a procedure in the bulk gravity
theory which deals with the UV divergence of the boundary theory using
the UV/IR relation of the AdS/CFT correspondence.  The bulk IR
regulator plays the role of the energy scale in the Wilsonian
renormalisation group (RG) of the boundary theory.  In the case of
spherical cut-off surface it is the radius of the sphere that is the
UV scale of the Wilsonian RG.  This makes it imperative to have a
proper boundary counter-term which correctly subtracts the background
contribution.

More concretely, the BK counter-terms, while
subtracting divergent pieces coming from the bulk geometry, do not
cancel the background contribution.  This is not relevant if we are
inserting operators at the boundary, but at any point away from the
boundary there are uncancelled bulk geometric contributions.  This
aspect becomes relevant when we have the RT surface dipping into the
bulk geometry.

In the context of AdS$_4$ geometry, it is possible to introduce
another counter-term, the Dirac-Born-Infeld (DBI) counter-term
\cite{Jatkar:2011ue}, whose large radius expansion contains the BK
counter-terms.  This counter-term has been shown to do full background
subtraction for pure AdS$_4$ geometry at any radial cut-off $r=r_C$
and in case of the AdS-Schwarzschild geometry, the on-shell action
evaluated on an $r=r_C$ surface correctly reproduces the black hole
entropy.  The DBI counter-term therefore seems to be a right candidate
to use for the computation of the holographic entanglement entropy.
However, before taking a plunge into this computation, it would be
desirable to carry out more checks on the applicability of the
Dirac-Born-Infeld counter-term.  {As an aside, it is
  worth mentioning that there exists an alternate formulation of
  counter-term, dubbed as the Kounterterm, in terms of the extrinsic
  curvature --- proposed by Olea \cite{Olea:2005gb,Olea:2006vd}.  This
  counter-term also does an equally good job and can be shown to agree
  with the DBI counter-term in the AdS$_{4}$ case by explicitly
  writing extrinsic curvature in terms of intrinsic one.  A detailed
  study of the Kounterterms in various
  dimensions \cite{Miskovic:2009bm} and its relation to the
  counter-terms written in terms of the intrinsic curvature is carried
  out in \cite{Anastasiou:2020zwc}.}

In this paper we will take up this task of consolidation of our
understanding of the counter-terms by studying the charged black holes
as well as local thermodynamics in AdS$_4$.  To be able to compute the
black hole entropy, it is essential to choose a correct ensemble.  For
example, in case of an electrically charged AdS-Reissner-Nordstr\"om
black hole, the grand canonical ensemble is appropriate, whereas for
the magnetically charged case we should use the canonical ensemble.
In section \ref{sec:dyonic-black-holes}, we will take up this problem
by considering the dyonic black holes in AdS$_4$, which would naturally be
defined in the mixed ensemble with electric charges in the grand
canonical and magnetic in the canonical ensemble.
However, it is quite inefficient to check validity of the counter-term
for each asymptotically AdS$_4$ background.  Instead a much more
convenient method is to device a technique which will ensure that the
counter-term does the desired job.  Studying local thermodynamics on
the cut-off surface precisely serves that purpose.  In section
\ref{sec:local-therm-at}, we show that the local thermodynamics is a
well defined notion on a smoothly deformed spherical cut-off surface.
We show that to the leading order in the deformation parameter, the
local thermodynamic quantities like charge and entropy function depend
on the deformation function.  This dependence, however, can be shown
to be a total derivative at the leading order and hence the integrated
quantities are not affected by the deformation.  This is a strong
result because although the deformation parameter is small compared to
the radius of the cut-off surface, the deformation function is
arbitrary and hence independence of thermodynamic quantities holds for
any arbitrary deformation of the cut-off surface.  That the DBI
counter-term performs complete background subtraction for the AdS$_4$
spacetime, in a way, suggests this counter-term would also be useful
for taking a flat-space limit of AdS, i.e., taking the cosmological
constant to zero.  We show in section \ref{sec:flat} that the
asymptotically flat limit is indeed well-defined and we recover the
correct thermodynamics.  In fact, the limiting value of the on-shell
action for a spherical cut-off surface matches that of the flat space
action with the usual Gibbons-Hawking prescription
\cite{Gibbons:1976ue} of background subtraction. We conclude with
summary of our results and some speculations.

\vspace{5mm}
\section{Dyonic Black Holes in AdS$_4$}\label{sec:dyonic-black-holes}

{Let us begin with the dyonic black hole in an
  asymptotically Anti-de Sitter (AdS) spacetime \cite{Lu:2013eoa,Lu:2013ura}}.  As
mentioned in the introduction, we will be using the mixed ensemble to
describe this black hole.  To compute thermodynamic quantities, it is
convenient to work with the Euclidean signature AdS space.

The bulk Euclidean action is the usual Einstein-Maxwell action in AdS
spacetime,
\be
I_{\mathrm{bulk}} = - \frac{1}{16\pi G} \int \dd[4]{x} \sqrt{g} \pqty{R - 2 \Lambda -   F_{\mu \nu} F^{\mu \nu} }, \label{actbulk}
\ee
where $F_{\mu \nu} = \del_\mu A_\nu - \del_\nu A_\mu$ is the Maxwell field strength and $\Lambda$ is the cosmological constant related to the AdS$_4$ length $L$ by,
\be 
\Lambda = - \frac{3}{L^2}.  \label{lamdef}
\ee
For a well-defined variational principle,  we have the Gibbons-Hawking-York boundary term,
\be 
I_{\mathrm{GHY}} = - \frac{1}{8\pi G} \int\limits_{\del} \dd[3]{x} \sqrt{\gamma} K,  \label{ighy}
\ee
where $\gamma_{\mu \nu}$ is the induced metric on the boundary $\del$ and $K$ is the trace of the extrinsic curvature tensor.  Instead of the BK counter-term,
{
\begin{equation}
  \label{eq:2}
  I_{\mathrm{BK}} = \frac{1}{8\pi G} \int\limits_\del \dd[3]{x} \sqrt{\gamma}\left[\frac{2}{L}+\frac{L}{2}R^{(3)} \right]\ ,
\end{equation}}
we shall use  the Dirac-Born-Infeld--like boundary counter-term \cite{Jatkar:2011ue},
\begin{equation}
I_{\mathrm{ct}} = \frac{L^2}{4\pi G} \int\limits_\del \dd[3]{x} \sqrt{-\det\pqty{ R^{(3)}_{\mu \nu} - \frac12  R^{(3)} \gamma_{\mu \nu} - \frac{1}{L^2} \gamma_{\mu \nu} } }, \label{ict}
\end{equation}
where the three-dimensional Ricci tensor and Ricci scalar are constructed out of the boundary metric $\gamma_{\mu \nu}$. 
In typical calculations,  one chooses a boundary hypersurface of large radius,  say $r_C$ and sends this cut-off surface to infinity, $r_C \to \infty$ at the end of the calculation.  

In particular, given a solution to the equations of motion, one can evaluate the on-shell action by substituting the solution in the action.  The free energy $\mathcal{F}$ of the system can then be written in terms of the on-shell action as,
\begin{equation}
\beta \mathcal{F} = I_{\mathrm{os}} \equiv \left(I_{\mathrm{bulk}}  + I_{\mathrm{GHY}} + I_{\mathrm{ct}}\right)_{\mathrm{os}},   \label{fren1}
\end{equation}
where $\beta$ is the inverse temperature.

Let us now write down the spherically symmetric dyonic solution to the Einstein-Maxwell equations of motion which follow from varying the action above.  The (Euclidean) metric is given by,
\begin{equation}
\dd{s}^2 = f(r)\dd{\tau}^2 + \frac{\dd{r}^2}{f(r)} + r^2 ( \dd{\theta}^2 + \sin^2 \theta \dd{\phi}^2),  \label{lineel}
\end{equation} 
where,
\begin{equation}
f(r) = 1 + \frac{r^2}{L^2} -  \frac{2GM}{r} + \frac{Q_e^2 + Q_m^2}{r^2}. \label{frbla}
\end{equation}
The gauge field which solves the Maxwell equations is given by,
\begin{equation}
A_\mu \dd{x^\mu} = - \mi Q_e \pqty{ \frac{1}{r_0} - \frac{1}{r} } \dd{\tau} + Q_m \cos \theta \dd{\phi}. \label{gaugfi}
\end{equation}
Here, $Q_e$ and $Q_m$ refer to the electric and magnetic charges
carried by the black hole.  Note that in our conventions $Q_e$ and
$Q_m$ have the dimensions of length.  We have to multiply by suitable
factors of $L$ and $G$ to make the charges dimensionless.  The factor
of $\mi$ in the expression \eqref{gaugfi} for the gauge field comes
from the analytic continuation $t = - \mi \tau$ of the solution in the
Lorentzian signature.  The black hole horizon is located at $r = r_0$
--- we have $f(r_0) = 0$ and $f(r) > 0$ for $r_0 < r < \infty$.  An
additive constant in the electric part of the gauge field in
\eqref{gaugfi} is suitably adjusted so that $A_\tau (r= r_0) = 0$,
which ensures that the gauge field is non-singular on the black hole
horizon.  {This solution is an Euclidean avatar of the
  solution presented in \cite{Lu:2013ura} where we have taken two
  harmonic functions $H_{1}$ and $H_{2}$ in \cite{Lu:2013ura} to be
  proportional to each other.  Equivalently, this is a solution with
  constant dilaton background.}

In the expression \eqref{frbla}, the quantity $M$ corresponds to the Arnowitt–Deser–Misner (ADM) mass of the black hole.   In all the expressions below,  we will write $M$ as a function of $Q_e$, $Q_m$ and $r_0$, so that we have,
\be
M = \frac{L^2 (Q_e^2+Q_m^2)+r_0^2(L^2 +r_0^2)}{2 G L^2 r_0}.  \label{admdef}
\ee
The black hole temperature as measured from asymptotic infinity is given by,
\baa
T =  \frac{r_0^2(L^2 +3 r_0^2) -L^2 (Q_e^2+Q_m^2)}{4 \pi  L^2 r_0^3}. \label{temdef}
\ea
The chemical potential corresponding to the electric charge is defined by,
\baa
\mu_e = \frac{1}{L} \mi A_\tau ( r \to \infty) = \frac{Q_e}{r_0 L}. \label{chempdef}
\ea

When we insert the solutions \eqref{lineel} and \eqref{gaugfi} in the action with a spherical boundary at $r = r_C$ and take the limit $r_C \to \infty$,  we obtain,
\be
\mathcal{F} = M -  T S  -  Q_e \mu_e \frac{L}{G}, \label{frendef1}
\ee
where $S = \pi r_0^2 / G$ is the Bekenstein-Hawking entropy.  We note that the on-shell action evaluates to a mixed ensemble --- it is the canonical ensemble with respect to the magnetic charge (in which the magnetic charge is held fixed), but the grand canonical ensemble with respect to the electric charge (in which the corresponding chemical potential is held fixed). We can find the entropy,  electric and magnetic charges by taking suitable derivatives of the free energy.  We will now show that we can reproduce the correct thermodynamics even at a finite cut-off $r = r_C$ and we can reproduce the correct values of the entropy and the dyonic charges from the free energy defined by \eqref{fren1},  where the ensemble continues to be the mixed one.  

We write down here only the final expressions.  When we evaluate the on-shell action with a finite cut-off, we have to be careful with quantities that are usually normalised with reference to the asymptotic infinity  $r =\infty$.  We have to have a non-trivial redshift factor which correctly accounts for the effects of the finite cut-off surface.  This can be interpreted as a normalisation of the temporal coordinate to maintain a unit speed of light on the boundary defined by $r=r_C$,  as was previously discussed in \cite{Jatkar:2011ue}.   For this cut-off surface,  the redshift factor $\rho$ is easily determined to be,
\baa
\rho = \frac{r_C}{L \sqrt{f(r_C)}} =  r_C^2 \sqrt{\frac{r_0}{(r_C-r_0) [ r_0 r_C \left(r_0 r_C+r_C^2+r_0^2\right)+ L^2 \left(r_0 r_C- Q_e^2 - Q_m^2\right) ]  }},  \label{rsfact1} 
\ea
which by construction approaches unity as $r_C \to \infty$.

This redshift factor must be taken into account while defining the free energy, temperature and chemical potential.  Therefore,  the appropriately renormalised free energy is given by,
\begin{equation}
\mathcal{F}_R = \frac{I_{\mathrm{os}}}{\beta} \rho, \label{rfrendef1}
\end{equation}
where the finite cut-off on-shell action is given by
\begin{align}
\frac{I_{\mathrm{os}}}{\beta} &= \frac{1}{GL^2} \sqrt{\frac{\left(r_C-r_0\right) \left(r_C^2+L^2\right) \left(r_0 r_C \left(r_0 r_C+r_C^2+r_0^2\right)+L^2 \left(r_0 r_C-Q_e^2 -Q_m^2\right)\right)}{r_0}} \nonumber \\
&\quad + \frac{r_C r_0 [3 L^2 r_0+r_0^3 -4 r_C (r_C^2+L^2)]+L^2 [ (5 r_C-4 r_0) Q_m^2 +  r_C Q_e^2]}{4 G L^2 r_0 r_C}.  \label{fcosa}
\end{align}
The  term on the first line above arises from the counter-term action \eqref{ict} and those on the second line arise from the bulk action \eqref{actbulk} and the GHY term \eqref{ighy}. The corrected temperature is given by,
\be
T_R = T \rho, \label{deftr}
\ee
(see eq.  \eqref{temdef})
and the chemical potential is evaluated from the value of the gauge field component $ \mi A_\tau$ on the cut-off surface. We have,
\be 
\mu_R = \frac{Q_e}{L} \pqty{ \frac{1}{r_0} - \frac{1}{r_C}  } \rho. \label{chempot}
\ee
Now note that in the mixed ensemble that we are working in,  the free energy is a function of $T_R$,  $\mu_R$ and $Q_m$.  Using the first law of thermodynamics,  we can determine the entropy $\hat{S}$,  electric charge $\hat{Q}_e$ and the magnetic chemical potential $\hat{\mu}_m$ by the relations,
\begin{align}
\hat{S} &= - \frac{\del \mathcal{F}_R}{\del T_R} \Bigg|_{\mu_R , Q_m},  \label{hats} \\
 \frac{L}{G} \hat{Q}_e &= - \frac{\del \mathcal{F}_R}{\del \mu_R} \Bigg|_{T_R , Q_m},  \label{hatqe}\\
\frac{L}{G}\hat{\mu}_m &= - \frac{\del \mathcal{F}_R}{\del Q_m} \Bigg|_{\mu_R ,  T_R }. \label{hatmum}
\end{align}
Note that we have defined these quantities with a hat because it is not {\it a priori} obvious that these quantities will be the same as those in the infinite cut-off theory.

All the thermodynamic quantities we have obtained above are functions of $r_0$,  $Q_m$,  $Q_e$ and $r_C$.  Since we keep the cut-off surface at $r=r_C$ fixed,  the effective variables are $(r_0 , Q_m , Q_e)$.  

For evaluating a quantity such as \eqref{hats},  we have to keep $\mu_R$ and $Q_m$ fixed.  This can be done as follows, we write,
\be 
\mu_R = \mu_R (r_0 , Q_m , Q_e). \label{mufunc1}
\ee 
A constant value of $\mu_R$ and $Q_m$ determines $Q_e$ as a function of $r_0$  --- which defines a special curve in the thermodynamic space.   Along this special trajectory,  $Q_e (r_0)$,  we have by an application of the chain rule,
\be 
Q'_e (r_0) =  -\frac{\pdv{\mu_R}{r_0}} {\pdv{\mu_R}{Q_e}}. \label{qprime1}
\ee
Therefore,  the entropy would be given by,
\be 
\hat{S} = - \frac{ \pdv{\mathcal{F}_R }{r_0} + \pdv{\mathcal{F}_R }{Q_e}  Q'_e (r_0) }{ \pdv{T_R }{r_0} + \pdv{T_R }{Q_e}  Q'_e (r_0)},  \label{entdef2}
\ee
where $\mathcal{F}_R$ and $T_R$ are treated as functions of $r_0$,  $Q_e$ and $Q_m$.   We will follow a similar strategy to determine the analogous partial derivatives.  Using the foregoing results, we obtain,
\be 
\hat{S} =  \frac{\pi r_0^2}{G}, \label{entbh}
\ee
which is precisely the expression for the Bekenstein-Hawking entropy! Let us see what we get for the other terms.  We obtain,
\be
\hat{Q}_e = Q_e,  \label{elchaeq}
\ee
and also,
\be 
\hat{\mu}_m = -\frac{Q_m}{L} \pqty{ \frac{1}{r_0} - \frac{1}{r_C}  } \rho, \label{magchemeq}
\ee
which is exactly what we expect from electric-magnetic duality (see eq.  \eqref{chempot}).

It is quite remarkable that in spite of the rather complicated expressions for the action \eqref{fcosa} and the redshift factor \eqref{rsfact1}, the correct answers are reproduced for any $r_C > r_0$ ---  various terms cancel to eventually yield the remarkably simple result.  It is worth mentioning here that we have gotten the correct expression for the electric charge and we used the unmodified magnetic charge even at finite cut-off because these quantities are independent of the radial cut-off on account of Gauss' law. In particular,  the expressions for the electric and magnetic charges are given by,
\begin{align}
Q_e  &\propto \int\limits_{r = r_C} * F,  \label{elchdef}\\
Q_m &\propto  \int\limits_{r = r_C} F,  \label{machdef}
\end{align}
where $F$ is the field-strength two-form and $*$ refers to the Hodge dual operation. These expressions are manifestly independent of $r_C$.  However,  this is not true of the conserved energy measured at $r=r_C$. Even if we set $Q_e = 0  = Q_m$, we find that,
\be 
E_R = \mathcal{F}_R + T_R S = \frac{r_C \left(r_C \sqrt{r_C^2+L^2}-\sqrt{r_C \left(r_C-r_0\right) \left(r_C^2+r_0 r_C+L^2+r_0^2\right)}\right)}{G L^2}. \label{cutoffen}
\ee
While this expression reduces to that of ADM mass \eqref{admdef} in the limit $r_C \to \infty$ (as it must),  this is not the ADM mass, even up to the redshift factor $\rho$. For now,  we can take this expression as the \emph{definition} of the mass in the absence of charges --- an independent verification would require a different calculation.  We can then work in the micro-canonical ensemble and we find using the definition \eqref{cutoffen} the relation,
\be 
T_R  \frac{\del S}{\del E_R} = 1.   \label{fltd1}
\ee

The reader would have noticed that the expression on the first line in \eqref{fcosa} (which comes from the counter-term) is manifestly invariant under an $\mathrm{O}(2)$ rotation of the electric and magnetic charges $(Q_e,  Q_m)$  but the expression on the second line of \eqref{fcosa} is not.  This is a reflection the fact that we are working in a mixed ensemble. If we want to work in the canonical ensemble with respect to both electric and magnetic charges,  we have to add a boundary term to the action \cite{Hawking:1995ap},
\begin{equation}
I_{\del \mathrm{GF}} = - \frac{1}{4\pi G} \int\limits_{\del} \sqrt{\gamma} n_\mu F^{\mu \nu} A_\nu,
\end{equation}
where $n_\mu$ is the (unit) one-form normal to the boundary. This contribution evaluates to,
\be
\frac{I_{\del \mathrm{GF}} }{\beta} = \frac{Q_e^2 (r_C-r_0) }{G r_0 r_C}. \label{idgf2}
\ee
By adding \eqref{idgf2} to the expression \eqref{fcosa},  we get an on-shell action that that is manifestly $\mathrm{O}(2)$-invariant and this corresponds to a free energy $\mathcal{F}_R$ for the canonical ensemble with respect to both electric and magnetic charges.  The free energy and temperature can then be regarded as functions of $r_0$, $Q_e$ and $Q_m$.  Now, we can keep $Q_e$ and $Q_m$ fixed and we obtain, as expected,
\begin{equation}
\hat{S} = - \eval{\pdv{\mathcal{F}_R}{T_R}}_{Q_e, Q_m} =  - \frac{ \pdv{\mathcal{F}_R }{r_0} }{ \pdv{T_R }{r_0} } = \frac{\pi r_0^2}{G}. \label{canentr}
\end{equation}
This shows that thermodynamics resulting from the Dirac-Born-Infeld--like counter-term \eqref{ict} is robust to different choices of ensemble.

{
\subsection{BK vs DBI Counter-term}
\label{sec:bk-vs-dbi}}

{After establishing the utility of the DBI counter-term,
  let us now briefly comment on the difference between the BK
  counter-terms and the DBI counter-terms.  We will also comment on
  what we mean by background subtraction.  The initial purpose of
  introducing the BK counter-terms was to ensure that the divergences,
  encountered when the radial cut-off in AdS is taken to infinity,  are
  removed.  This ensures finiteness of, say,  the computation of Witten diagrams --- which allows us to define a proper AdS/CFT dictionary
  so that bulk and boundary computations could be compared and/or
  translated into each other.  On the other hand, in the formulation
  of holographic renormalisation group (RG),  the radial coordinate of
  the AdS space behaves like the RG scale.  In the Wilsonian
  formulation,  under the RG flow,  we would be writing an effective
  field theory at some finite energy scale.  This, in the holographic
  RG picture,  would correspond to carrying out computations at a
  finite radial cut-off.  In this case, there are no divergences but
  if we try to compute the on-shell action then we get a series of
  terms, some of which growing as the cut-off is taken to infinity and
  others vanishing in the same limit.  Computation of the on-shell
  action means we are only looking at the contribution of the AdS
  background to the Brown-York energy momentum tensor.  Ideally, we
  would like the Brown-York tensor to contain information about
  fluctuations only and not contain any information about the AdS
  background.  In this sense it is desirable to have a counter-term
  which removes all contribution of the background geometry and
  carries information about fluctuations about the background.  Any
  counter-term which implements this is said to be carrying out the
  complete background subtraction.  As we take the cut-off to
  infinity, the terms in the on-shell action that are growing have
  dominant contribution.  Those set terms are taken care of by the BK
  counter-term but besides those there are terms, which do not
  contribute as the cut-off was taken to infinity --- that do contribute
  at a finite cut-off.  This contribution to the Brown-York
  energy-momentum tensor at $r=r_{C}$ is not removed by the BK
  counter-term.  For illustration, the value of on-shell action at a
  finite cut-off, $r=r_{C}$, in the pure AdS case with
  $S^{2}\times S^{1}$ boundary and with the BK counter-term \eqref{eq:2} is
\begin{equation}
  \label{eq:1}
I_{\mathrm{os}} =  \frac{L^2 \beta }{8 G r_{C}}+\mathcal{O} \left(\frac{1}{r_{C}^5}\right)\ ,
\end{equation}
where $L$ is the AdS radius.  On the other hand, the DBI counter-term
cancels even these $r_C$-dependent contributions for pure AdS space with
various boundary topologies including $S^{2}\times S^{1}$  and $S^3$,  and
therefore the on-shell action for all these boundary topologies of
AdS$_{4}$  are $r_C$-independent.  As a result, up to a possible overall constant term,  the Brown-York tensor
contains contributions coming from fluctuations only.  For the
AdS-Schwarzschild geometry, the DBI counter-term correctly reproduces
the black hole thermodynamics at any cut-off surface.}

{If we do a derivative expansion of the DBI
  counter-term then first couple of terms are in exact agreement with
  the BK counter-term, and the rest of the terms ensure that at a
  finite radius the classical background contribution is subtracted, up to a universal term, if it is present.
  It is therefore more appropriate to say that the BK counter-term and
  the DBI counter-term are simply two different subtraction schemes
  and the advantage of the DBI term is it simplifies certain
  computations, {\it e.g.}, black hole thermodynamics in case of
  asymptotically AdS black hole solutions as we just saw.  Later we will comment on its advantage in the computation
  of the entanglement entropy using the Ryu-Takayanagi
  prescription \cite{JatkarMoitra2}.}

\vspace{5mm}

\section{Local Thermodynamics at Finite Cut-Off}\label{sec:local-therm-at}

In this section, we will argue that on a finite cut-off surface of arbitrary shape,  the local Lagrangian gives rise to a sensible \emph{local} notion of thermodynamics. 

To be concrete,  we consider not a spherical cut-off surface $r=r_C$,  but a surface defined by the relation,
\be 
r = r_C + \epsilon Y (\theta), \label{wigsud}
\ee
where $\epsilon$ is taken to be a small parameter ($\epsilon \ll r_C$) and $Y(\theta)$ is an arbitrary but smooth function on the interval $\theta \in [0, \pi]$.  We would have to impose some additional constraints on $Y(\theta)$ near the ``poles'' $\theta = 0$ and $\theta = \pi$ to avoid conical singularities.  In all our calculations below,  we shall be evaluating all quantities to linear order in $\epsilon$.  To this order,  the induced metric on  the cut-off surface reads,
\be
\dd{s}_\Sigma^2 =  \bqty{ f(r_C) + \epsilon f'(r_C) Y(\theta)  } \dd{\tau}^2 + \bqty{ r_C^2 + 2 \epsilon r_C Y(\theta) } \pqty{ \dd{\theta}^2 + \sin^2 \theta \dd{\phi}^2  },  \label{indmet}
\ee
and so the conical singularity is not seen at this order.  We actually get a more stringent condition  on the smallness of $\epsilon$ from the $\tau \tau$ component of the metric above, 
\baa
\epsilon \ll \frac{f(r_C)}{f'(r_C)}. \label{epscon}
\ea
When we were considering spherical cut-off surfaces as in the previous section,  the cut-off surface at $r=r_C$ could be taken arbitrarily close to the black hole horizon at $r=r_0$.  However,  when we have a non-spherical surface such as \eqref{wigsud},  the condition \eqref{epscon} implies that we should also have $\epsilon \ll r_C - r_0$.

The red-shift factor on the cut-off surface will now be different at different values of $\theta$ and we shall have a local red-shift factor given by,
\be 
\rho(\theta) = \frac{r_C}{L \sqrt{f\left(r_C\right)}}+\frac{\epsilon  Y(\theta ) \left(2 f\left(r_C\right)-r_C f'\left(r_C\right)\right)}{2 L f\left(r_C\right){}^{3/2}}+\mathcal{O}(\epsilon ^2).  \label{locrs}
\ee

We define the local free energy on each boundary point by using the previously considered action but only doing the radial bulk integral (the remaining terms are evaluated on the boundary \eqref{wigsud}),
\begin{align}
\mathfrak{f} (\theta) = \rho(\theta) &\Bigg[- \frac{1}{16\pi G} \int\limits_{r_0}^{r_C + \epsilon Y(\theta)} \dd{r} \sqrt{g} (R - 2\Lambda - F^2) - \frac{1}{8\pi G} \sqrt{\gamma} K \nonumber\\
&\quad + \frac{L^2}{4\pi G} \sqrt{-\det\pqty{ R^{(3)}_{\mu \nu} - \frac12 R^{(3)} \gamma_{\mu \nu} - \frac{1}{L^2} \gamma_{\mu \nu} }} \Bigg].  \label{locfth}
\end{align}
We can define the local temperature $\mathfrak{t} (\theta)$ and chemical potential $\mathfrak{m} (\theta) $ with the same relations as before,  with suitable adjustments,
\begin{align}
\mathfrak{t} (\theta) &= T \rho (\theta),  \label{loctem} \\\
\mathfrak{m} (\theta) &=  \frac{Q_e}{L} \pqty{ \frac{1}{r_0} - \frac{1}{r_C} + \frac{\epsilon Y(\theta)}{r_C^2}  + \mathcal{O} (\epsilon^2) } \rho (\theta). \label{loctth}
\end{align}

Taking these to be the local definitions of free energy,  temperature and chemical potential,  we can find the local thermodynamic conjugate quantities (entropy and charge) by taking suitable derivatives,
\begin{align}
\mathfrak{s} (\theta) &= - \frac{\del \mathfrak{f} (\theta)}{\del \mathfrak{t} (\theta)},  \label{locsd}\\
\frac{L}{G} \mathfrak{q} (\theta) &= - \frac{\del \mathfrak{f} (\theta)}{\del \mathfrak{m} (\theta)}. \label{locqd}
\end{align}

We find for the entropy an answer of the form,
\be
\mathfrak{s} (\theta) =  \frac14 \frac{r_0^2}{G} \sin \theta + \epsilon \bqty{ Y'(\theta) \cos \theta + Y''(\theta) \sin \theta }    F_1 + \mathcal{O} (\epsilon^2) . \label{sloc1}
\ee
In the second term above, $F_1$ is a function of $R$, $L$, $r_0$ and the $Q$'s whose precise form is not important at the moment. The reader would have noticed that the $\theta$-dependent parts can be written as a total $\theta$-derivative of the function $\big( \sin \theta \, Y'(\theta) \big)$. Therefore,  we obtain,  to this order of approximation,
\be 
\int_0^{2\pi} \dd{\phi} \int_0^\pi \dd{\theta} \mathfrak{s} (\theta) = \frac{\pi r_0^2}{G} . \label{sint1}
\ee
Therefore,  the local entropy function,  when integrated over the cut-off surface gives precisely the Bekenstein-Hawking entropy,  with the additional piece in \eqref{sloc1} integrating to zero.\footnote{The quantity $\mathfrak{s} (\theta)$ is essentially an entropy density function on a unit $S^2$,  with the integration measure factored in.  The first term in \eqref{sloc1} corresponds to a constant entropy density on the unit $S^2$.} This also gives us the constraint that $Y(\theta)$ should be regular near $\theta = 0$ and $\theta = \pi$ so that the boundary contributions vanish.

By similar manipulations,  we find an exactly analogous answer for the local charge function,
\be 
\mathfrak{q} (\theta) =  \frac{1}{4\pi} Q_e \sin \theta + \epsilon \bqty{ Y'(\theta) \cos \theta + Y''(\theta) \sin \theta }    F_2 + \mathcal{O} (\epsilon^2) . \label{sqoc1}
\ee
Therefore,  this function, integrated, gives the correct charge. 

The results above suggest that local quantities which can be defined
from the Lagrangian density are sensible thermodynamic quantities.  It
is worth noting that the $\mathcal{O} (\epsilon)$ term in both
\eqref{sloc1} and \eqref{sqoc1} is a total $\theta$-derivative relies
on several aspects, including the correct red-shift factor
$\rho(\theta)$.  We must emphasise, however, that there could be
inherent ambiguities in such local thermodynamic quantities.  For
example, the additional term in \eqref{sqoc1} cannot be interpreted as
coming from the integrand of \eqref{elchdef}.  What essentially
happens here is that definition \eqref{locqd} \emph{effectively}
corresponds to adding a local charge density which integrates to zero.
Similar comments apply for the entropy density as well.

We considered the scenario where $Y$ was only a function of $\theta$. We could consider a general function of $Y(\theta, \phi)$ and expand it in terms of spherical harmonics.  The calculation is similar.  We find,  in this case,  that the additional $\mathcal{O} (\epsilon)$ contribution has a $\theta, \phi$-dependence of the form,
\be 
\pdv{\theta} \pqty{ \sin \theta  \pdv{Y}{\theta}  } +  \frac{1}{\sin \theta} \pdv[2]{Y}{\phi}. \label{ylap}
\ee
This would also integrate to zero. However,  we would now require to $Y$ to vanish at least linearly around $\theta = 0$ and $\theta  = \pi$.

\vspace{5mm}
\section{Thermodynamics for Asymptotically Flat Spacetime}\label{sec:flat}

One of the most appealing features of the counter-term \eqref{ict} is that it achieves a complete background subtraction in asymptotically Anti-de Sitter spacetimes,  as mentioned previously.  This fact,  in tandem with the previous results,  motivates us to examine the possibility that we could obtain the correct thermodynamics for asymptotically flat spacetimes in the limit of a vanishing cosmological constant $\Lambda\to 0$ (i.e.,  in the limit of an infinite AdS radius $L \to \infty$).  

It is worth pointing out that the flat-space limit of the on-shell action \eqref{fren1} is well-defined is a non-trivial feature of the counter-term \eqref{ict}.  On the other hand,  the on-shell action with the BK counter-term,  in contrast,  diverges as $L\to \infty$ for a fixed $r_C$.  For example,  for pure thermal AdS (i.e.,  the solution \eqref{lineel} with $M = 0  = Q_{e,m}$) with a cut-off at $r=r_C$,  we have, 
\begin{equation}
\frac{I_{\mathrm{BK}} }{\beta} = - \frac{r_C}{G} \pqty{ 1 + \frac{r_C^2}{L^2}  }  + \frac{\sqrt{L^2 + r_C^2}}{2G}   \pqty{ 1 + \frac{2r_C^2}{L^2}  },   \label{ibalak}
\end{equation}
which diverges linearly for large $L$ and becomes ill-defined.  The on-shell action \eqref{fren1}, on the other hand,  vanishes for pure AdS$_{4}$ spacetime for any value of $L$ and therefore also has the same value in the flat-space limit.

However,  naively taking the $L \to \infty$ limit of the answers in the previous sections gives the wrong answer and we have to be a bit more careful,  especially with the red-shift factor $\rho$. Let us first take the $L \to \infty$ limit in the evaluation of the on-shell action,  \eqref{fcosa}.  This limit is well-defined and we obtain,
\begin{align}
\frac{(I_{\mathrm{os}})_{\mathrm{flat}}}{\beta} &= \lim_{L \to \infty} \frac{I_{\mathrm{os}}}{\beta} \nonumber \\
&= \frac{1}{G}\sqrt{\frac{\left(r_C-r_0\right) \left(r_0 r_C-Q_e^2-Q_m^2\right)}{r_0}} + \frac{r_C Q_e^2+\left(5 r_C-4 r_0\right) Q_m^2}{4 G r_0 r_C} + \frac{3 r_0-4 r_C}{4 G}.  \label{iflat}
\end{align}
But the $L \to \infty$ limit of the red-shift factor $\rho$,  eq.  \eqref{rsfact1},  gives a vanishing result for all values of $r_C$. Therefore, we cannot take this limit naively --- otherwise we would arrive at a manifestly wrong result that the free energy, temperature and chemical potential vanish for asymptotically flat spacetime.  By the usual elementary considerations,  the red-shift factor at a radial location $r=r_C$ for asymptotically flat spacetime is given by the inverse square-root of the Schwarzschild factor $f(r)$,
\begin{equation}
\rho_{ \mathrm{flat} } = \frac{1}{\sqrt{f(r_C)}} = \pqty{ 1 - \frac{2GM}{r_c} + \frac{Q_e^2 + Q_m^2}{r_c^2}  }^{-1/2}.  \label{rhoflat}
\end{equation}
We also have to be careful with the definition of the charge and chemical potential while taking the flat-space limit.  In these definitions,  we make the replacement $L \to \sqrt{G}$ for dimensional consistency.  For example, the electric chemical potential reads,
\begin{equation}
\mu_{R} = \frac{Q_e}{\sqrt{G}} \pqty{ \frac{1}{r_0} - \frac{1}{r_C}  } \rho_{\mathrm{flat}} .  \label{muflat}
\end{equation}
With these modifications,  we find suitable individual definitions for the finite cut-off free energy, temperature and chemical potential.  Taking the derivatives as before,  we find the correct expressions for entropy,  electric charge and magnetic chemical potential.

Note that the expression for the on-shell action in the flat space limit is exactly what one gets from the method of background subtraction due to Gibbons and Hawking \cite{Gibbons:1976ue},
\begin{equation}
I = - \frac{1}{16\pi G} \int \dd[4]{x} \sqrt{g} \pqty{ R - F^2 } - \frac{1}{8\pi G} \int\limits_{\partial}\dd[3]{x} \sqrt{\gamma} \pqty{  K - K[\eta] },  \label{gibha}
\end{equation}
where $K$ corresponds to the usual trace of the extrinsic curvature in the spacetime described by the metric $g_{\mu \nu}$, while $K[\eta]$ corresponds to the extrinsic curvature of the same surface embedded in a flat spacetime with the metric $\eta_{\mu \nu}$.  This prescription is rather ad-hoc in nature.   However,  from the point of view of  asymptotically flat spacetime arising as an infinite-radius limit of AdS spacetime,  this background subtraction arises naturally.   The background-subtracting boundary counter-term  \eqref{ict} is constructed only from the boundary data in the spacetime with metric $g_{\mu \nu}$ and does not refer to any ``external'' reference spacetime\footnote{The extrinsic curvature $K[\eta]$ can be derived by using the pullback metric on the spherical cut-off surface.}.

Before ending this section,  let us mention an important aspect vis-à-vis the infinite-$L$ limit. While it is true that, for spherical cut-off surfaces,  the on-shell action \eqref{fren1} is well-defined as $L\to \infty$,  it turns out that there is a divergence when we consider a surface of the form  \eqref{wigsud}.  This is evident in the behaviour of    the function $F_1$ that appears in \eqref{sloc1} which, for large $L$, grows quadratically in $L$,   $F_1 \sim L^2$.  Therefore,  naively, for a non-constant function $Y(\theta)$,  the local entropy function $\mathfrak{s} (\theta)$ is not well-defined.  This can be remedied by redefining the infinitesimal parameter $\epsilon$ suitably so that even as we take the limit $L \to \infty$,  $\epsilon L^2 / R^3$ continues to remain small.  We could even start in flat spacetime with the  action  \eqref{gibha} and consider such arbitrary surfaces as before. In this case,  the AdS length scale $L$ does not appear at all and so we need not worry about such divergences.  Quite reassuringly,  the local notion of thermodynamics as formulated in section 
\ref{sec:local-therm-at} makes sense in this case and we recover the correct entropy and charge by integrating over the $\theta, \phi$ directions.

\vspace{5mm}

\section{Discussion}\label{sec:discuss}

It was known that the Dirac-Born-Infeld--like counter-term effects
exact background cancellation in the pure AdS$_4$ case and in the
AdS-Schwarzschild case it correctly reproduced black hole
thermodynamics at any spherical cut-off surface.  We have shown here
that these results continue to hold for charged dyonic black holes in
AdS spacetime and flat spacetime as well.  To establish this result,
one needs to make an appropriate choice of ensemble.  Ideally it
should be possible to extend this procedure to Kerr-AdS black holes as
well except that finding equal acceleration surface for a rotating
black hole is trickier than what appears at the first glance.  We
nevertheless employed same procedure with a spherical cut-off surface
whose radius is larger that the radius of the ergosphere and executed
it numerically to find the results agreeing with the Kerr black hole
thermodynamics within a few parts in $10^5$.  We have not presented
these results here for want of obtaining exact match with the Kerr
thermodynamics.  The analytic computation is significantly more
complicated in the Kerr-AdS case, however, a judicious choice of
cut-off surface may lead to exact results.  We intend to report on it
in future.  We have instead worked out another aspect related to the
thermodynamics of asymptotically AdS geometries, namely, we showed
that one can define a notion of local thermodynamic functions on
spherical cut-off surfaces deformed by a small parameter multiplied by
an arbitrary function of coordinates orthogonal to the radial
coordinate.  We showed that at the leading order in the deformation
parameter the deviation from the spherical geometry does not affect
thermodynamic quantities.  The local thermodynamic functions do depend
on the deformation but that dependence can be shown to be a total
derivative term and as a result it does not contribute to
thermodynamic quantities obtained by integrating the local functions
over the compact surface.  It is worth re-emphasising that this result
is independent of choice of the deformation function.  We also showed
that this procedure works equally well for asymptotically flat
spacetimes by considering the dyonic black hole solution.
{This was demonstrated in section \ref{sec:flat}.  It
  is natural to speculate if this counter-term has any role to play in
  the flat space holography.  In particular, it would be interesting
  to see if the DBI counter-term has any role to play in the Bondi–van der Burg–Metzner–Sachs (BMS)
  symmetry on the null infinity of the asymptotically flat spacetime.
  As natural extension one can think of extending this to
  asymptotically de Sitter spacetime.  In this context it is worth pointing
  out that cancellation of divergence gave two solutions in
  \cite{Jatkar:2011ue}.  The DBI counter-term used here is one of the
  solutions and the other solution is
  \begin{equation}
    I_{\mathrm{ct}} =- \frac{\mathrm{i}L^{2}}{4\pi G}\sqrt{-\det(R_{\mu\nu}^{(3)}
          -\frac{1}{6}R^{(3)}\gamma_{\mu\nu}
        +\frac{1}{L^{2}}\gamma_{\mu\nu} ) }\ .
  \end{equation}
  It was speculated in \cite{Jatkar:2011ue} that the latter may be
  relevant for the de Sitter case.  It would be interesting to
  explicitly check which solution is relevant in the de Sitter case.
  However, this exploration is beyond the scope of this paper.}

Our original motivation for reanalysing this counter-term was to use
it in the computation of the holographic entanglement entropy.  The
arbitrariness of the deformation function implies that small
deformation of the surface does not affect computation of the physical
quantities.  This will be a useful criterion in the computation of the
area of the minimal surface bounding two entangling regions in the
boundary theory.  {One of the interesting outcomes of
  the DBI counter-term in computation of the entanglement
  entropy in the global AdS$_{4}$ is that the contribution of the
  Ryu-Takayanagi surface could vanish identically for an appropriate region.  A direct implication
  of this is the entanglement entropy in this case would have contribution only from the
  fluctuation across the Ryu-Takayanagi surface and none from the
  background geometry.  We will report on it in more detail soon \cite{JatkarMoitra2}.}

\vspace{5mm}

\noindent{\bf Acknowledgments:} We would like to thank S. Ashok and
A. Laddha for discussion.  DPJ would like to acknowledge support from
the ICTP through the Associates Programme (2022-2027) as well as
support from SERB grant CRG/2018/002835.  UM thanks the ICTP for support.

 \bibliographystyle{JHEP}
 
 \bibliography{refs}

\end{document}